\documentclass[journal=jacsat,manuscript=communication,etalmode=truncate,keywords=true]{achemso}
\setkeys{acs}{articletitle=true,maxauthors=10,etalmode=truncate,keywords=true}
\usepackage{amsmath,amssymb,units,txfonts}
\usepackage{amsfonts}

\usepackage{graphicx}
\usepackage{MnSymbol}
\usepackage{helvet,wasysym}
\usepackage[bookmarks=true,colorlinks=true,urlcolor=blue,linkcolor=blue,citecolor=blue]{hyperref} 
\usepackage[usenames,dvipsnames]{color}
\usepackage{colortbl,cuted,bm}
\usepackage{multirow}
\usepackage{xr}

\usepackage[utf8x]{inputenc}
\usepackage{xfrac}

\definecolor{JPCCBlue}{RGB}{34,80,169}
\definecolor{JCPCGreen}{RGB}{11,122,64}
\definecolor{StyleColor}{RGB}{34,80,169}
\definecolor{abstractcolor}{RGB}{255,243,201}
\makeatletter\newenvironment{abstractbox}{%
   \begin{lrbox}{\@tempboxa}\begin{minipage}{0.988\textwidth}}{\end{minipage}\end{lrbox}%
   \colorbox{abstractcolor}{\usebox{\@tempboxa}}
}\makeatother

\renewcommand*\authorfont{\flushleft}
\renewcommand*\affilfont{\mdseries\flushleft\textrm}
\renewcommand*\titlesize{\Large}
\renewcommand*\titlefont{\flushleft\sf\bfseries}
\def\refname{\sffamily\color{StyleColor}\Large$\blacksquare$\normalsize\ REFERENCES}
\usepackage{titlesec}
\titleformat{\section}{\bfseries\sffamily\color{StyleColor}}{\thesection.~}{0pt}{}
\titleformat{\subsection}[runin]{\bfseries\sffamily\normalsize}{\indent\thesubsection.~}{0pt}{}[.]
\titlespacing{\subsection}{0pt}{0pt}{*1}
\titleformat{\subsubsection}{\bfseries\sffamily\normalsize}{\thethesubsection.~}{0pt}{}
\titlespacing{\subsubsection}{0pt}{0pt}{*0}

\newcommand{\new}[1]{#1}
\newcommand{\old}[1]{}

\title{Using \textit{G}$_{\text{0}}$\textit{W}$_{\text{0}}$ Level Alignment to Identify Catechol's Structure on TiO$_{\text{2}}$(110) }
\author{Duncan J.~Mowbray}
\email{duncan.mowbray@gmail.com}
\affiliation[UPV/EHU]{\newline\footnotemark[2]{\ } Nano-Bio Spectroscopy Group and ETSF Scientific Development Center, Departamento de F{\'{\i}}sica de Materiales, 
 Universidad del Pa{\'{\i}}s Vasco UPV/EHU and DIPC, E-20018 San Sebasti\'{a}n, Spain}
\author{Annapaola Migani}
\email{annapaola.migani@icn2.cat}
\affiliation[ICN2CSIC]{\newline\footnotemark[3]{\ } Catalan Institute of Nanoscience and Nanotechnology (ICN2), CSIC and The Barcelona Institute of Science and Technology, Campus UAB, Bellaterra, E-08193 Barcelona, Spain}

\begin{document}
\maketitle

\begin{strip}
\vspace{-1.cm}

\noindent{\color{StyleColor}{\rule{\textwidth}{0.5pt}}}
\begin{abstractbox}
\begin{tabular*}{17cm}{b{11.5cm}r}
\noindent\textbf{\color{StyleColor}{ABSTRACT:}}
We perform state-of-the-art calculations for a prototypical dye sensitized solar cell: catechol on rutile TiO$_2$(110). Catechol is often used as an anchoring group for larger more complex  organic and inorganic dyes on TiO$_2$ and forms a type II heterojunctions on TiO$_2$(110). In particular, we compare quasiparticle (QP) $G_0W_0$ with hybrid exchange correlation functional (HSE) density functional theory (DFT) calculations for the catechol-rutile TiO$_2$(110) interface.  In so doing, we provide a theoretical interpretation of ultraviolet photoemission spectroscopy (UPS) and inverse photoemission spectroscopy (IPES) experiments for this prototypical system. 
Specifically, we demonstrate that the position, presence, and intensity of peaks associated with catechol's HOMO, intermolecular OH--O bonds, and  interfacial hydrogen bonds to the surface bridging O atoms (O$_{\textit{br}}$H--C and O$_{\textit{br}}$H--O) may be used to fingerprint deprotonation of catechol's OH anchoring groups.  Furthermore, our results suggest deprotonation of these groups, while being nearly isoenergetic at high coverages, may significantly increase the photovoltaic efficiency of catechol--TiO$_2$(110) interfaces.
&\includegraphics[width=5.1cm]{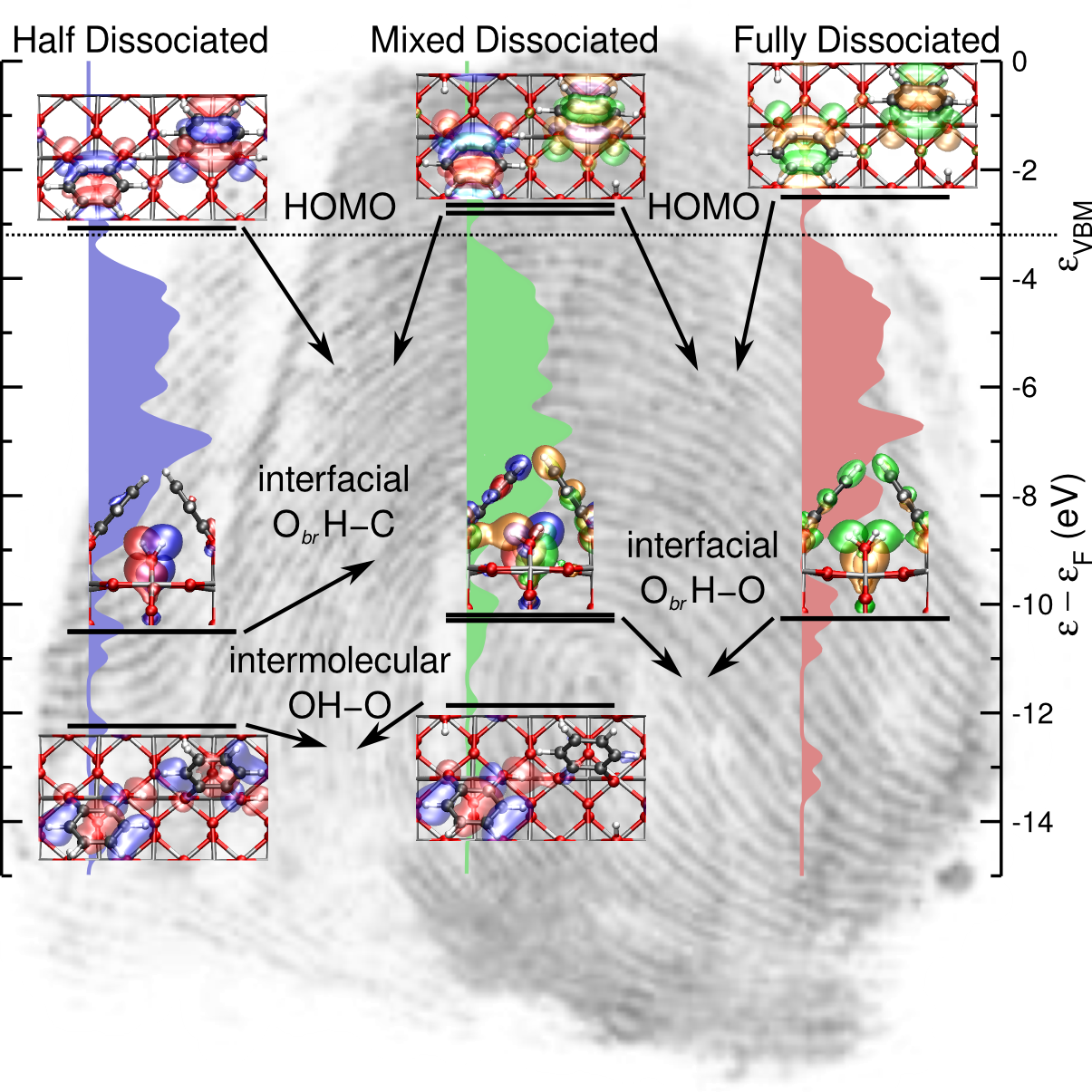}
\end{tabular*}
\end{abstractbox}
\noindent{\color{StyleColor}{\rule{\textwidth}{0.5pt}}}
\end{strip}

\def\bigfirstletter#1#2{{\noindent
    \setbox0\hbox{{\color{StyleColor}{\Huge #1}}}\setbox1\hbox{#2}\setbox2\hbox{(}%
    \count0=\ht0\advance\count0 by\dp0\count1\baselineskip
    \advance\count0 by-\ht1\advance\count0 by\ht2
    \dimen1=.5ex\advance\count0 by\dimen1\divide\count0 by\count1
    \advance\count0 by1\dimen0\wd0
    \advance\dimen0 by.25em\dimen1=\ht0\advance\dimen1 by-\ht1
    \global\hangindent\dimen0\global\hangafter-\count0
    \hskip-\dimen0\setbox0\hbox to\dimen0{\raise-\dimen1\box0\hss}%
    \dp0=0in\ht0=0in\box0}#2}

\section{INTRODUCTION}

Catechol on TiO$_2$(110) is a prototypical system for modelling industrially relevant dye sensitized solar cells \cite{Diebold,DieboldCatecholScience,CalzolariJPhysChemC2012,RisplendiPhysChemChemPhys2013,MaromJPCLett2014,PrezhdoJPhysChemB2005,PrezdhoAnnRevPhysChem2006}. This is for two reasons. (1) Catechol is often used as an anchoring group \cite{CalzolariJACS2011} for larger more complex organic and inorganic dyes. (2) Catechol forms type II heterojunctions \cite{CalzolariJACS2011}, where the dye's  highest occupied molecular orbital (HOMO) is a gap state, while the lowest unoccupied molecular orbital (LUMO) is above the conduction band minimum (CBM) of the substrate, e.g., TiO$_2$.  The level alignment for this system has previously been studied experimentally via ultraviolet photoemission spectroscopy (UPS) \cite{Rangan20104829,Diebold} and inverse photoemission spectroscopy (IPES) \cite{Rangan20104829}, and thus provides an excellent benchmark type II system to be studied theoretically using many-body quasiparticle (QP) techniques, such as $G_0W_0$\cite{GW,AngelGWReview,KresseG0W0}.  
For such type II systems, the HOMO--LUMO and HOMO--CBM separations determine the onset of the \old{adsorption}\new{absorption} spectra.  In such photovoltaic devices, the dye's role is to reduce the onset of the \old{adsorption}\new{absorption} spectra to sub-band gap energies, and maximize the overlap with the solar spectrum\cite{MaromJPCLett2014,DeAngelisDSSC}.

The exact structure of the catechol--rutile TiO$_2$(110) interface is difficult to identify and control experimentally.  This is because for hydroxylated molecules such as catechol, a complex network of interfacial and intermolecular bonds is formed upon adsorption.  Moreover, catechol's anchoring OH groups can be either fully dissociated, partially dissociated, or intact, with each catechol overlayer having similar adsorption energies.  Still, dissociation plays a significant role in determining the level alignment\cite{Diebold,ThorntonH2ODissTiO2110} and, hence, the photovoltaic efficiency\cite{DeAngelisDSSC} of the interface.  

Generally, the HOMO moves to higher energy upon deprotonation for hydroxylated molecules, e.g., CH$_3$OH on TiO$_2$(110)\cite{OurJACS} and H$_2$O on rutile TiO$_2$(110)\cite{MiganiH2OJCTC2014} and anatase TiO$_2$(101)\cite{SunH2OAnatase}.
For this reason, the HOMO's energy provides a fingerprint of the interface's structure.   This is augmented by the presence of distinct identifiable levels associated with intermolecular and interfacial OH--O hydrogen bonds.  Specifically, the hydrogenated bridging O atoms (HO$_{\textit{br}}$) resulting from interfacial deprotonation of the anchoring groups are consistently at $\sim 10$~eV below the Fermi level $\varepsilon_F$ on TiO$_2$(110)\cite{ThorntonH2ODissTiO2110,MiganiH2OJCTC2014,DeSegovia,Krischok}.

To describe both highly hybridized \cite{MiganiH2OJCTC2014} and localized \cite{OurJACS,MiganiInvited} molecular interfacial levels, one requires a correct description of the anisotropic screening, i.e., electron-electron interaction,  at the interface. This is clearly seen from the level alignment of H$_2$O on TiO$_2$(110)\cite{MiganiH2OJCTC2014}. 

On the one hand, H$_2$O's 1b$_1$ level becomes highly hybridized with the substrate upon adsorption on TiO$_2$(110) \cite{MiganiH2OJCTC2014}.  The interfacial level alignment of these highly hybridized levels with the substrate VBM is already described by density functional theory (DFT). This is because the screening of these levels is essentially the dielectric constant of the substrate. 
At the $G_0W_0$ level,  the DFT Kohn-Sham (KS) eigenenergies are shifted by $\Delta$, the difference between the QP self-energy ($\Sigma = i G_0W_0$) and the xc potential ($V_{\textrm{xc}}$), up to a normalization factor $Z$, i.e., $\Delta \equiv Z(\Sigma - V_{\textrm{xc}})$\cite{KresseG0W0}.    For example, $\Delta \approx 0.1$  and $0.7$ eV for all levels of the TiO$_2$(110) substrate from $G_0W_0$ based on DFT calculations employing a generalized gradient approximation (PBE)\cite{PBE} and a hybrid range separated (HSE)\cite{HSE06} exchange and correlation (xc)-functional, respectively\cite{MiganiLong}.  

 On the other hand, H$_2$O's 3a$_1$ and 1b$_2$ levels remain mostly localized on the molecule upon adsorption on TiO$_2$(110) \cite{MiganiH2OJCTC2014}.  The alignment of these localized levels, whose screening is significantly different from that of the substrate, is poorly described at the DFT level.  Not even HSE\cite{HSE06} provides an accurate description of the interfacial level alignment for localized levels, e.g., H$_2$O's 3a$_1$ and 1b$_2$ and CH$_3$OH's a$^{_{''}}$ on TiO$_2$(110)\cite{MiganiH2OJCTC2014,MiganiLong}.  
This is because HSE DFT calculations effectively perform a constant static screening of the exchange term, i.e., the fraction of Hartree-Fock exact-exchange included $\alpha = 0.25$ is effectively an inverse dielectric constant $\alpha \approx \varepsilon_\infty^{-1}$\cite{Marques}.  For this reason, HSE performs well for systems with a homogeneous screening and  $\varepsilon_\infty\sim 4$\cite{Marques,MiganiLong,MiganiH2OJCTC2014}.

To correctly account for differences in screening between the molecular layer and substrate, one may use many-body QP techniques, such as $G_0W_0$\cite{AngelGWReview}.  In such methods, the spatial dependence of the screening is included explicitly.  In fact, for occupied levels, $\Delta$ is linearly dependent on the fraction of the wave function's density within the molecular layer $f_{\textit{mol}}$\cite{OurJACS,MiganiLong,SchefflerGWsurfaces}.  This means by just rigidly shifting all the levels, one cannot describe the alignment of occupied levels with significant density outside the substrate.

For type I interfaces, i.e., H$_2$O or CH$_3$OH on TiO$_2$(110)\cite{MiganiH2OJCTC2014,OurJACS,MiganiLong,MiganiInvited}, we have previously demonstrated that $G_0W_0$ provides an accurate alignment for both localized and highly hybridized levels.   In each case, $G_0W_0$ shifts the localized levels to stronger binding, into quantitative agreement with UPS experiments.  

In this study, we compare the projected density of states (PDOS) onto catechol obtained from QP PBE $G_0W_0$ and HSE DFT calculations with the measured UPS and IPES spectra for the catechol--rutile TiO$_2$(110) interface.  In so doing, this study provides a complete state-of-the-art computational description of the simplest experimentally relevant type II interface.  
Based on our analysis of the PDOS, we are able to \old{determine the precise}\new{suggest the most likely} structure of the catechol overlayer measured in UPS experiments.  In fact, \old{we find}\new{our results suggest} the degree of catechol dissociation \old{differs markedly}\new{may differ} with the experimental conditions employed\cite{Rangan20104829,Diebold}.  Our results suggest fully deprotonating the anchor groups of the overlayer should lead to an increased efficiency of the photovoltaic device.

\section{METHODOLOGY}\label{Sect:Methodology}

Our $G_0W_0$ calculations\cite{GW,AngelGWReview,KresseG0W0} have been performed using \textsc{vasp} within the projector augmented wave (PAW) scheme \cite{kresse1999}.   The $G_0W_0$ calculations are based on KS wave functions and eigenenergies from DFT obtained using PBE\cite{PBE}. 
DFT calculations employing the HSE06 variant\cite{HSE06} of the HSE xc-functional have been carried out for comparison with PBE $G_0W_0$ calculations.

In the $G_0W_0$ approach, the contribution to the KS eigenvalues from the xc-potential $V_{xc}$ is replaced by the QP self energy $\Sigma = i G_0 W_0$ in a single step, where $G_0$ is the Green's function and $W_0$ is the screening \cite{GW} based on the KS wave functions and eigenvalues \cite{AngelGWReview}.  The dielectric function is obtained from linear response time-dependent (TD) DFT within the random phase approximation (RPA), including local field effects \cite{KresseG0W0}.  From $G_0W_0$ one obtains first-order QP corrections $\Delta$ to the KS eigenvalues, but retains the KS wave functions.

  The geometries have been fully relaxed using the PBE\cite{PBE} xc-functional, with all forces $\lesssim$ 0.02 eV/\AA.  We employ a plane-wave energy cutoff of 445 eV, an electronic temperature $k_B T\approx0.2$ eV with all energies extrapolated to $T\rightarrow 0$ K, and a PAW pseudopotential for Ti which includes the 3$s^2$ and 3$p^6$ semi-core levels.  The calculations have been performed spin unpolarized.  All unit cells contain a four layer TiO$_2$(110) slab, employ the measured lattice parameters of bulk rutile TiO$_2$  ($a=4.5941$ \AA, $c=2.958$ \AA)\cite{TiO2LatticeParameters}, and include at least 27 \AA\ of vacuum between repeated images. In each case, equivalent catechol overlayers are adsorbed on both sides of the slab.  We employ $\Gamma$ centered $k$-point meshes with densities $\Delta k < 0.25$~\AA$^{-1}$, approximately 9\sfrac{1}{4} unoccupied bands per atom, i.e.\ including all levels up to 30~eV above the VBM, an energy cutoff of 80 eV for the number of \textbf{G}-vectors, and a sampling of 80 frequency points for the dielectric function.   The $G_0W_0$ parameters are consistent with those previously  used  for describing both rutile and anatase TiO$_2$ bulk, rutile TiO$_2$(110) and anatase TiO$_2$(101) clean surfaces, and their interfaces\cite{OurJACS,MiganiLong,MiganiH2OJCTC2014,SunH2OAnatase}.  These parameters have been shown to provide accurate descriptions of bulk rutile and anatase optical absorption spectra, and both clean surface and interfacial level alignment\cite{OurJACS,MiganiLong,MiganiH2OJCTC2014,SunH2OAnatase}.  

The adsorption energy $E_{\textit{ads}}$ of catechol on \old{the}\new{Ti coordinately unsaturated (}Ti$_{\textit{cus}}$\new{)} sites of a TiO$_2$(110) surface is given by
\begin{equation}
E_{\textit{ads}} \approx \frac{E[n\textrm{Catechol}+\textrm{TiO}_2\textrm{(110)}] - E[\textrm{TiO}_2\textrm{(110)}]}{n}- E[\textrm{Catechol}],
\end{equation}
where $n$ is the number of adsorbed catechol molecules in the supercell,
and $E[n\textrm{Catechol}+\textrm{TiO}_2\textrm{(110)}]$,
$E[\textrm{TiO}_2\textrm{(110)}]$, and $E[\textrm{Catechol}]$ are the total
energies of the covered and clean surfaces and gas phase catechol
molecule, respectively.  For catechol in the gas phase, we find the most stable conformation has an intramolecular hydrogen bond, i.e., $C_S$ symmetry.

Scanning tunneling microscopy (STM) simulations have been performed using the Tersoff-Hamann approximation\cite{TersoffHamann}.  In this approach, the current $I$ at a position $\textbf{r}$ is given by
\begin{equation}
I(\textbf{r}) = \new{C \int_{\varepsilon_F}^{\varepsilon_F+U} \rho(\textbf{r}, \varepsilon) d\varepsilon \approx CU \rho(\textbf{r}, \varepsilon_F)} \approx CU \rho(\textbf{r}\new{, \varepsilon_F+U}),
\end{equation}
where $C$ is a prefactor which depends on the DOS, surface work function, and tip radius. $U$ is the potential of the sample relative to the tip in the experiment, i.e., the applied potential relative to the experimental Fermi level $\varepsilon_F$.  $\rho(\textbf{r}\new{, \varepsilon})$ is the local DOS, given by
\begin{equation}
\rho(\textbf{r}\new{,\varepsilon}) = \new{\sum_{\new{n}\textbf{k}} \delta(\varepsilon_{n\textbf{k}} - \varepsilon)  \frac{|\psi_{n\textbf{k}}(\textbf{r})|^2}{N_{\textbf{k}}}}\new{\approx} \sum_{\new{n}\textbf{k}} \exp\left[-\left(\frac{\varepsilon_{n\textbf{k}} - \old{U}\new{\varepsilon}}{k_BT}\right)^2\right]  \frac{ |\psi_{n\textbf{k}}(\textbf{r})|^2}{N_{\textbf{k}}},
\end{equation}
where $\varepsilon_{n\textbf{k}}$ is the $G_0W_0$ eigenvalue and $\psi_{n\textbf{k}}$ is the KS wave function of level $n$ at $k$-point $\textbf{k}$, $k_BT \approx 0.2$ eV is the electronic temperature of the calculation, and $N_{\textbf{k}}$ is the weight of $k$-point $\textbf{k}$.  \new{To emphasize any dependence of $I(\textbf{r})$ on the applied bias $U$, we have used throughout $I(\textbf{r}) \approx C U\rho(\textbf{r}, \varepsilon_F+U)$.  It should be noted, however, that similar results are obtained by integrating $\rho$ over the bias window $[\varepsilon_F, \varepsilon_F + U]$ (see Figure S1 in Supporting Information).  Herein,} $\rho$ is plotted at \new{an isosurface value of }$5\times 10^{-8}\ e/$\AA$^3$.  This is somewhat greater than its maximum far from the surface, i.e., 
\begin{equation}
\new{\min_z\left(\max_{x,y}\rho(x,y,\old{0}\new{z};\varepsilon_F+U)\right)} < 5\times 10^{-8}\new{\ e/\textrm{\AA}^3}.
\end{equation}
\new{This ensures $\rho$ is defined at this isosurface value throughout the surface plane.}

Experimental spectra are typically referred to the Fermi level, $\varepsilon_F$, which is pinned $\sim0.1$~eV below the CBM for mildly reduced TiO$_{\text{2}}$ \cite{TiO2Fermi,TiO2FermiAono,TiO2FermiYamakata}.  Using  the electronic band gap for rutile TiO$_{\text{2}}$ of $3.3 \pm0.5$~eV obtained from electron spectroscopy measurements \cite{TiO2BandGap}, the experimental VBM energy relative to the Fermi level is $\varepsilon_{\textrm{VBM}} \approx 0.1 - 3.3 \approx -3.2$~eV\cite{MiganiLong}.  Since the VBM is the most reliable theoretical energy reference\cite{MiganiLong}, we subtract $\varepsilon_{\textrm{VBM}} \approx -3.2$~eV from the measured UPS and IPES spectra to align with the calculated $G_0W_0$ DOS and PDOS, and vice versa.

We align the DOS and PDOS with respect to the deepest Ti semi-core level (3s$^2$).   This allows a direct comparison between spectra for half (\sfrac{1}{2}D \sfrac{1}{2}D), mixed (D \sfrac{1}{2}D), and fully (D D) dissociated catechol overlayers.  In each case, the highest occupied levels belong to the catechol overlayer.  As we are interested in seeing the difference between the HOMO position and the TiO$_2$(110) VBM for each catechol overlayer, the highest occupied level is not a good reference. 

Moreover, the catechol anchor groups form Ti$_\textit{cus}$--O bonds, making it impractical to separate the O 2p$_\pi$ surface levels from those of the molecule.
  This makes it difficult to identify a highest occupied level with purely surface contributions, which can be associated with the clean surface's VBM.  For this reason, one should use the Ti levels as reference, e.g., occupied semi-core Ti 3s$^2$ or unoccupied Ti 3d CBM levels.   Here, we align with respect to the deepest semi-core Ti 3s$^2$ levels.  In this way we obtain a consistent alignment relative to the surface levels  for each configurations.  We find this is effectively equivalent to aligning relative to the TiO$_2$ CBM, i.e., Ti 3$d$ levels. 

Using the semi-core levels, we remove differences in work function between surfaces, which would be present if the vacuum level were used as a reference.  Here, we take the energy of the VBM for the clean surface $\varepsilon_{\mathrm{VBM}}^{\mathrm{clean}}$ relative to its deepest Ti 3s$^2$ semi-core level $\varepsilon_{\mathrm{Ti}3s^2}^{\mathrm{clean}}$ as our final reference relative to the catechol interface's deepest Ti 3s$^2$ level $\varepsilon_{\mathrm{Ti}3s^2}$.  More precisely, $\varepsilon_{\mathrm{VBM}} = \varepsilon_{\mathrm{VBM}}^{\textit{clean}} - \varepsilon_{\mathrm{Ti}3s^2}^{\textit{clean}} + \varepsilon_{\mathrm{Ti}3s^2}$, in Figures \ref{fgr:OrbitalSpectra} and \ref{CatecholPDOS:fgr}.  

However, for reduced TiO$_{2-x}$(110), where Ti 3d levels are occupied, all the Ti levels are consequently upshifted compared to stoichiometric TiO$_2$(110).  This makes the Ti levels a poor reference for comparison between such systems.  For this reason, we use the VBM as an energy reference for 1 ML H@O$_{\textit{br}}$ on TiO$_2$(110), as it is a type I interface.

\section{RESULTS AND DISCUSSION}

Catechol consists of a benzene ring with two adjacent anchoring OH groups.  It has been previously shown, both theoretically and experimentally, that catechol adsorbs on \old{coordinately unsaturated (}\new{the }Ti$_{\textit{cus}}$\old{)} sites of the rutile TiO$_2$(110) surface via the OH anchor groups in a bidentate configuration.

At low coverage, catechol preferentially adsorbs upright ($\theta \approx 86^\circ$) on the surface, parallel to the [001] $c$-axis \cite{DieboldCatecholScience}, with both anchoring groups deprotonated, i.e., fully dissociated, with an accompanying charge of $-0.4e$ transferred to the nearest O$_{\textit{br}}$ atom of the TiO$_2$(110) surface.    As the coverage increases ($\sim\sfrac{2}{3}$ ML), catechol tilts ($\theta \sim 67^\circ$) toward the surface, with two interfacial O$_{\textit{br}}$H--O bonds (\textit{cf.} Table~\ref{TableS1}).  

\begin{table}
\vspace{-19mm}
\noindent{\color{StyleColor}{\rule{\columnwidth}{1.0pt}}}
\vspace{14mm}
\caption{\textrm{\bf Adsorption Energies \textit{E}$_{\textit{ads}}$ in Electronvolts per Molecule for \sfrac{1}{2} ML 1$\times$4, \sfrac{2}{3} ML 1$\times$3, and 1 ML 1${\times}$4 \new{Intact (I), Half (\sfrac{1}{2}D) and Fully (D) Dissociated }Catechol Overlayers on TiO$_{\text{2}}$(110) with Tilting Angle $\boldsymbol{\theta}$ in Degrees and Number of Interfacial (OH--O) and Intermolecular (OH--O$_{\textit{br}}$,O$_{\textit{br}}$H--C, and O$_{\textit{br}}$H--O) Bonds per Unit Cell }}\label{TableS1}
\tiny
\begin{tabular}{c@{}ccccccc}
\multicolumn{8}{>{\columncolor[gray]{0.9}}c}{ }\\[-1mm]
\multicolumn{2}{>{\columncolor[gray]{.9}}c}{structure} & 
\multicolumn{1}{>{\columncolor[gray]{.9}}c}{$\theta$} & 
\multicolumn{1}{>{\columncolor[gray]{.9}}c}{OH--O} & 
\multicolumn{1}{>{\columncolor[gray]{.9}}c}{OH--O$_{\textit{br}}$} & 
\multicolumn{1}{>{\columncolor[gray]{.9}}c}{O$_{\textit{br}}$H--C} & 
\multicolumn{1}{>{\columncolor[gray]{.9}}c}{O$_{\textit{br}}$H--O} &
\multicolumn{1}{>{\columncolor[gray]{.9}}c}{$E_{\textit{ads}}$}\\
\multicolumn{1}{>{\columncolor[gray]{0.9}}c}{(ML)} &
\multicolumn{1}{>{\columncolor[gray]{0.9}}c}{ } &
\multicolumn{1}{>{\columncolor[gray]{0.9}}c}{($^\circ$)} &
\multicolumn{4}{>{\columncolor[gray]{0.9}}c}{(bonds/unit cell) }&
\multicolumn{1}{>{\columncolor[gray]{0.9}}c}{(eV)}
\\[0.5mm]



%
\multirow{1}{*}{\sfrac{1}{2}}& \multirow{1}{*}{D} & 
 \multicolumn{1}{c}{86} & 
0 & 0 & 0 & 2 &
-0.792 \\\hline

\multirow{7}{*}{\sfrac{2}{3}}& \old{\multirow{2}{*}{I}}\new{I} & 
 \multicolumn{1}{c}{87}  & 
0 & 0 & 0 & 0 &
-0.249\\
&\new{I}& 
 \multicolumn{1}{c}{61}  & 
0 & 2 & 0 & 0 &
-0.595\\

& \old{\multirow{2}{*}{\sfrac{1}{2}D}}\new{\sfrac{1}{2}D} & 
 \multicolumn{1}{c}{80} & 
0 & 1 & 0 & 1 & 
-0.620 \\
&\new{\sfrac{1}{2}D}& 
 \multicolumn{1}{c}{63} & 
0 & 1 & 0 & 1 &
-0.676 \\

& \old{\multirow{3}{*}{D}}\new{D} & 
 \multicolumn{1}{c}{78} &  
0 & 0 & 2 & 0 &
-0.676 \\
& \new{D} & 
 \multicolumn{1}{c}{87} & 
0 & 0 & 0 & 2 &
-0.705 \\
& \new{D} & 
 \multicolumn{1}{c}{67} & 
0 & 0 & 0 & 2 &
-0.748\\\hline

\multirow{6}{*}{1}& \new{\sfrac{1}{2}D \sfrac{1}{2}D} & 
 56, 49 & 
2 & 0 & 2 & 0 &
-0.614\\
& \sfrac{1}{2}D \sfrac{1}{2}D & 
 55, 48 & 
1 & 1 & 0 & 2 &
-0.600\\
& \new{\sfrac{1}{2}D \sfrac{1}{2}D} &
 54, 46 & 
2 & 0 & 0 & 2 &
-0.598\\

&\old{\multirow{2}{*}{D \sfrac{1}{2}D}}\new{D \sfrac{1}{2}D} & 
 56, 48 & 
1 & 0 & 1 & 2 &
-0.685\\
& \new{D \sfrac{1}{2}D} & 
 56, 48 & 
1 & 0 & 2 & 1 &
-0.653\\

&D D & 
 56, 49 & 
0 & 0 & 2 & 2 &
-0.652\\
\end{tabular}
\noindent{\color{StyleColor}{\rule{\columnwidth}{1.0pt}}}
\end{table}

\begin{figure}
\includegraphics[width=\columnwidth]{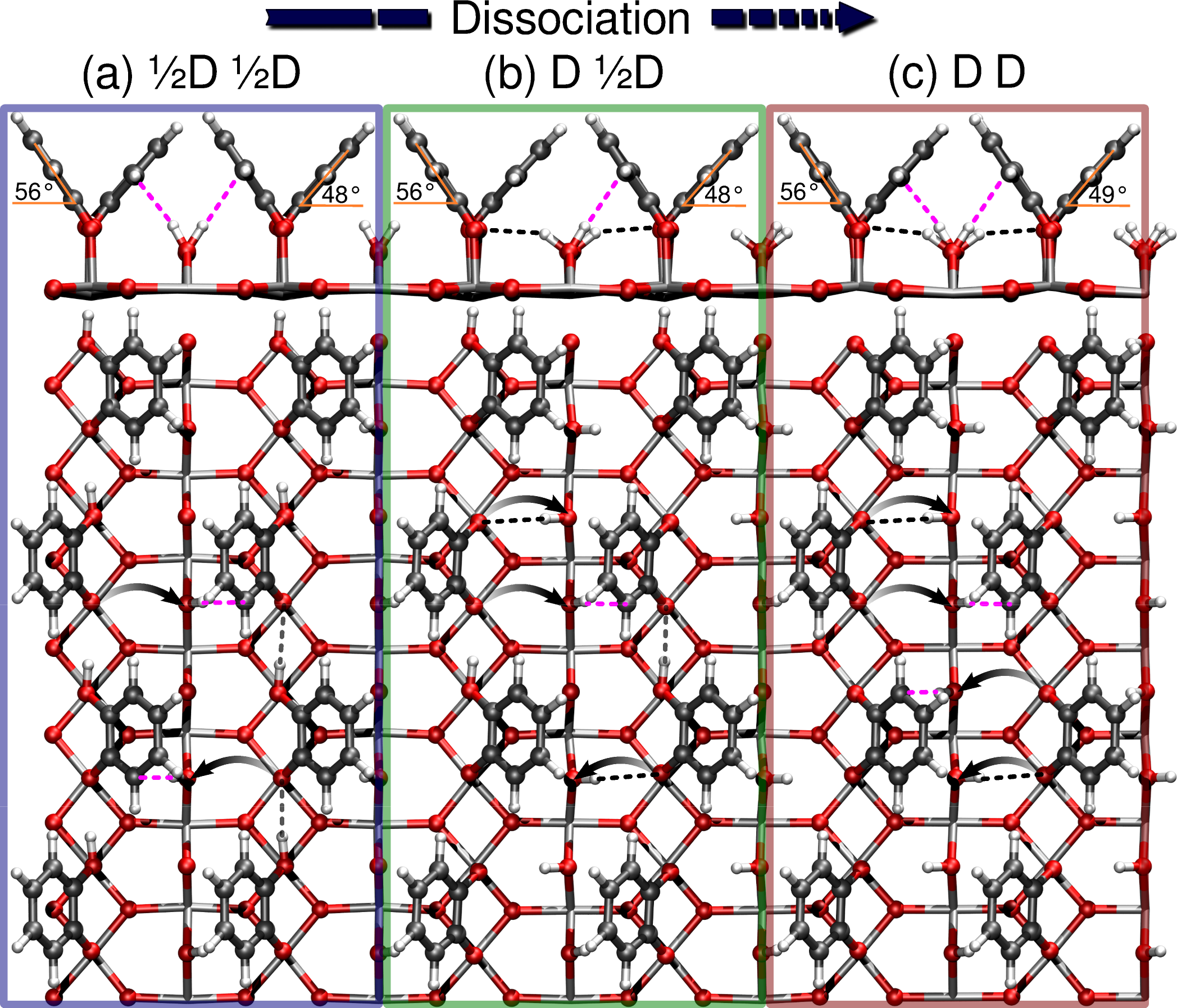}
\caption{Schematics of 1 ML catechol adsorbed (a) half (\sfrac{1}{2}D \sfrac{1}{2}D, blue), (b) mixed (D \sfrac{1}{2}D, green), and (c) fully (D D, red) dissociated on coordinately unsaturated Ti sites (Ti$_{\textit{cus}}$) of TiO$_2$(110).  Charge transfer of $\sim-0.4 e$ accompanying deprotonation is represented by arrows, while intermolecular OH--O (gray) and interfacial O$_{\textit{br}}$H--O (black) and O$_{\textit{br}}$H--C (magenta) hydrogen bonds are denoted by dotted lines.  The angle between catechol's benzene ring and the surface plane, $\theta$, is shown above.
}\label{Structures:fgr}
\noindent{\color{StyleColor}{\rule{\columnwidth}{1pt}}}
\end{figure}

At high coverage (1 ML), where catechol molecules adsorb on every Ti$_{\textit{cus}}$ site, they are forced to tilt in alternating directions due to steric hindrance.  This gives rise to the $1\times4$ catechol overlayers seen via STM\cite{Diebold} and shown in Figure~\ref{Structures:fgr}.  
Further, at this coverage, besides interfacial hydrogen bonds to the surface HO$_{\textit{br}}$, the overlayer, when not fully dissociated, is stabilized by intermolecular hydrogen bonds between neighboring catechol molecules.  For this reason, the half, mixed, and fully dissociated catechol adsorption energies are all within 0.1 eV, i.e., the accuracy of DFT (\textit{cf.} Table~\ref{TableS1}).  These results have also been reproduced to within 25 meV using the real space PAW DFT code \textsc{gpaw}\cite{gpaw1,gpaw2}.

Overall, we find catechol binds more weakly to the surface as the coverage increases. This is attributable to steric hindrance, especially for a 1 ML coverage. Furthermore, the binding energy at \sfrac{2}{3} ML coverage is significantly stronger for deprotonated anchoring groups and tilted catechol molecules (\textit{cf.} Table~\ref{TableS1}).  This is consistent with previous DFT studies of catechol on TiO$_2$(110)\cite{RisplendiPhysChemChemPhys2013}.

The most stable 1 ML half, mixed, and fully dissociated catechol overlayers listed in Table~\ref{TableS1} are shown in Figure~\ref{Structures:fgr}.  The
1 ML \sfrac{1}{2}D \sfrac{1}{2}D structure (Figure~\ref{Structures:fgr}(a)) has two intermolecular OH--O (gray dashed lines) and two interfacial O$_{\textit{br}}$H--C bonds (magenta dashed lines), the 1 ML D \sfrac{1}{2}D structure (Figure~\ref{Structures:fgr}(b)) has one intermolecular OH--O, one interfacial O$_{\textit{br}}$H--C, and two interfacial O$_{\textit{br}}$H--O bonds (black dashed lines), while the 1 ML D D structure  (Figure~\ref{Structures:fgr}(c)) has two interfacial O$_{\textit{br}}$H--C and two interfacial O$_{\textit{br}}$H--O bonds.  In each case, interfacial deprotonation of the anchoring OH groups is accompanied by a charge transfer of $-0.4e$ to the nearest O$_{\textit{br}}$ atom, as depicted schematically in Figure~\ref{Structures:fgr}.
Such intermolecular and interfacial hydrogen bonding combinations have also been reported for the 1 ML methanol--TiO$_2$(110) interface \cite{Jin,OurJACS}.  
Figure~\ref{Structures:fgr} shows that the two tilting directions are inequivalent, with adjacent catechols  tilted by $\theta \sim 56^\circ$ and $48^\circ$.  
For the O$_{\textit{br}}$H--O interfacial hydrogen bonds, catechol tilts away from the O$_{\textit{br}}$H moiety, while for the O$_{\textit{br}}$H--C interfacial hydrogen bonds, catechol tilts towards the O$_{\textit{br}}$H moiety.  For the latter, the O$_{\textit{br}}$H--C bond is mostly HO$_{\textit{br}}$ $\sigma$ in character, with a minor C 2p$_\pi$ contribution from the neighboring C atoms.  

\begin{figure*}[!t]
\includegraphics[width=\textwidth]{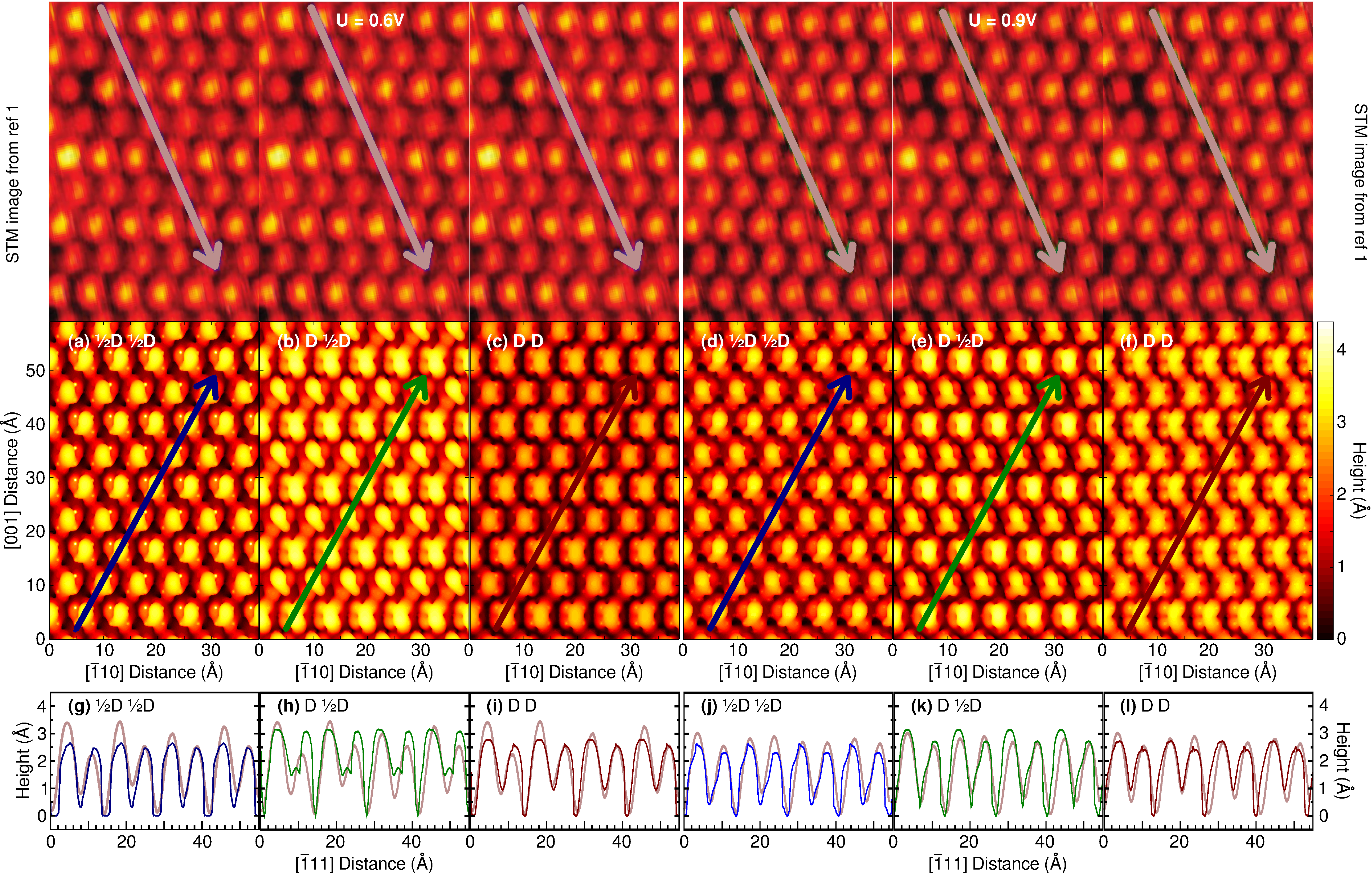}
\caption{Constant current \textbf{(a-f)} STM and \textbf{(g--l)} line scans at (left panels) U = 0.6 V  and (right panels) U = 0.9 V from ref\old{.}~\citenum{Diebold} (upper panels and brown thick lines) and calculated LDOS \new{(lower panels and blue/green/red thin lines) }for \textbf{(a,d,g,j)} half (\sfrac{1}{2}D \sfrac{1}{2}D, blue), \textbf{(b,e,h,k)} mixed (D \sfrac{1}{2}D, green), and \textbf{(c,f,i,l)} fully (D D, red) dissociated catechol overlayers on TiO$_2$(110) at $5\times10^{-8}\ e/$\AA$^{3}$. \new{Heights in \AA\ are relative to the measured or calculated minimum.}
}\label{STM:fgr}
\noindent{\color{StyleColor}{\rule{\textwidth}{1pt}}}
\end{figure*}

STM images of the $1\times 4$ catechol overlayer\cite{Diebold}, and simulated images for the \sfrac{1}{2}D \sfrac{1}{2}D, D \sfrac{1}{2}D, and D D structures are shown in Figure~\ref{STM:fgr}.  In order to reproduce the line scan's minima, i.e., effective height, it is necessary to perform the line scan along the [$\overline{1}11$] direction, as indicated by arrows in Figure~\ref{STM:fgr}(a--f).  This is particularly important for the D D structure, where a line scan along the [$\overline{1}{1}\overline{1}$] direction would not cross the computed STM minima.   Comparing the measured and computed line scans along the [$\overline{1}11$] direction, we thus conclude that the measured catechol overlayer is the mirror image of the computed structures shown in Figure~\ref{Structures:fgr}.

Overall, the \new{average heights $h$ from }line scans and simulated STM images agree qualitatively for all three computed structures at both $U=0.6$\new{~V ($h \approx 2.79$ \AA, $h_{\textrm{exp}} \approx 2.87$ \AA)} and $\new{U= }0.9$~V\new{ ($h \approx 2.67$ \AA, $h_{\textrm{exp}} \approx 2.75$ \AA)}. \new{However, the difference in height between neighbouring molecules $\Delta h$ is underestimated at $U=0.6$~V ($\Delta h \approx 0.14$~\AA, $\Delta h_{\textrm{exp}} \approx 0.81$~\AA), but agrees qualitatively at $U=0.9$~V ($\Delta h \approx 0.32$~\AA, $\Delta h_{\textrm{exp}} \approx 0.22$~\AA).}   \old{This}\new{Altogether, this} indicates that the structure of the unoccupied levels at these biases are rather insensitive to the deprotonation of catechol's anchoring OH groups.  This is not surprising, considering the structural similarity of the benzene ring orientation for each type of overlayer shown in Figures \ref{Structures:fgr} and \ref{STM:fgr}.


In summary, STM provides direct information as to the relative orientation of catechol on the surface which forms the $1\times4$ overlayer on TiO$_2$(110).  However, STM lacks direct information about deprotonation of the OH anchor groups. Such information is important, as it determines the relative level alignment of the molecule's HOMO with the substrate's VBM.  
To obtain direct information about the level alignment, one must compare the $G_0W_0$ PDOS with UPS and IPES spectra.  

\begin{figure*}[!t]
\includegraphics[width=\textwidth]{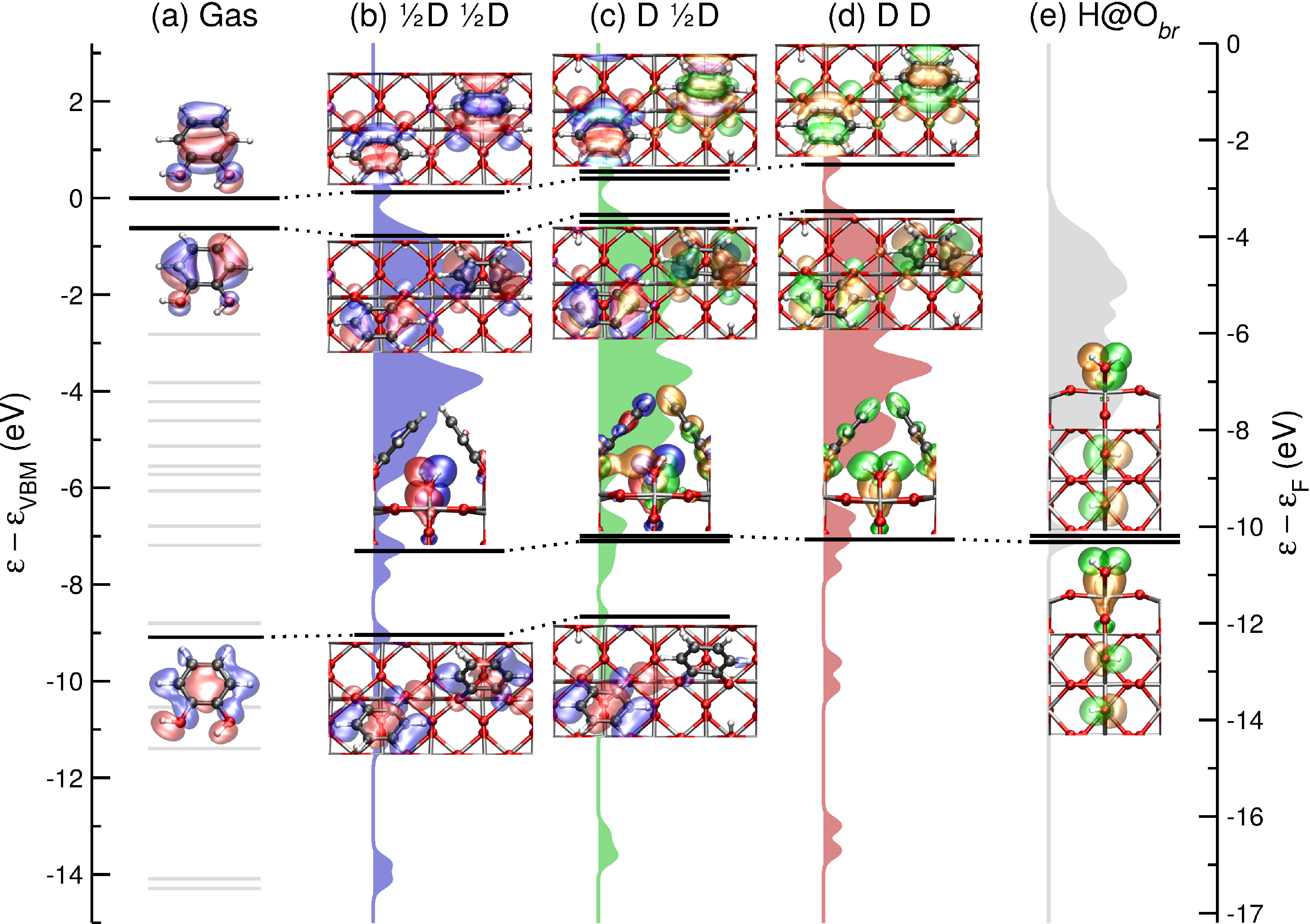}
\caption{HOMO, HOMO$-1$, OH--O, and HO$_{\textit{br}}$ DFT orbitals and $G_0W_0$ eigenvalues and DOS for catechol (a) in the gas phase (grey lines), 1 ML (b) half (\sfrac{1}{2}D \sfrac{1}{2}D, blue), (c) mixed (D \sfrac{1}{2}D, green), or (d) fully dissociated (D D, red), and  (e) H@O$_{\textit{br}}$ (grey) on TiO$_2$(110).  
(b,c,d) Colors are used to differentiate between \sfrac{1}{2}D (red/blue) and D (orange/green) catechol orbitals.  
 Energies are relative to the VBM ($\varepsilon_{\textrm{VBM}}$) of clean TiO$_2$(110) (left) or the experimental Fermi level (right).  
}\label{fgr:OrbitalSpectra}
\noindent{\color{StyleColor}{\rule{\textwidth}{1pt}}}
\end{figure*}

In Figure~\ref{fgr:OrbitalSpectra}, we identify features of the $G_0W_0$ PDOS which may be used to fingerprint the catechol overlayer's structure.  
These features are the HOMO, HOMO$-1$, interfacial HO$_{\textit{br}}$, and intermolecular HO--O levels associated with the half (blue/red) and fully (green/orange) dissociated surface catechol species.  Dotted lines connect energy levels associated with the half or fully dissociated catechol species.  

Although catechol's gas phase spectrum contains many additional occupied $\pi$ and $\sigma$ skeletal levels, shown in gray in Figure~\ref{fgr:OrbitalSpectra}(a), these levels are rather insensitive to deprotonation of the OH anchor groups.  For this reason, they are ineffective for distinguishing between \sfrac{1}{2}D \sfrac{1}{2}D, D \sfrac{1}{2}D, and D D catechol overlayers on TiO$_2$(110).

As a reference, the HOMO and HOMO$-1$ levels of gas phase catechol are depicted in Figure~\ref{fgr:OrbitalSpectra}(a), aligned relative to the molecule's HOMO.  
Figure~\ref{fgr:OrbitalSpectra}(b), (c), and (d) show that the HOMO and HOMO$-1$ are pinned to each other, and shift up in energy with deprotonation of adsorbed catechol.  This deprotonation of the OH anchor group induces a charge transfer of $\sim-0.4e$ to the substrate.  As charge is removed from the molecule, the HOMO and HOMO$-1$ are destabilized.  This effect is even more pronounced at the $G_0W_0$ level, as the molecule's ability to screen the HOMO and HOMO$-1$ levels is also reduced as charge is transferred to the substrate\cite{OurJACS}.  Consequently, the energy separation between HOMO and VBM may be used to distinguish between half, mixed, and fully dissociated catechol overlayers on TiO$_2$(110).  


Another fingerprint of dissociated catechol is the presence of HO$_{\textit{br}}$ surface levels at $\sim 10$ eV below the experimental Fermi level.   The HO$_{\textit{br}}$ level is a general feature of all the interfaces formed from rutile TiO$_2$(110) and hydroxylated molecules, e.g., H$_2$O and CH$_3$OH\cite{MiganiH2OJCTC2014}.  
As a reference, we show in Figure \ref{fgr:OrbitalSpectra}(e) both in and out of phase HO$_{\textit{br}}$ $\sigma$ levels for 1 ML H@O$_{\textit{br}}$, on TiO$_2$(110)\cite{MiganiH2OJCTC2014}. This structure is equivalent to \sfrac{1}{2} ML dissociated H$_2$O adsorbed on bridging O vacancies (H$_2$O@O$_{\textit{br}}^{\textit{vac}}$) of a reduced TiO$_{2-\textrm{\sfrac{1}{4}}}$(110) surface \cite{MiganiH2OJCTC2014}.  
 In Figure~\ref{fgr:OrbitalSpectra}(c,d) we show HO$_{\textit{br}}$ levels associated with adjacent O$_{\textit{br}}$H--O bonds, while for (b), we show O$_{\textit{br}}$H--C levels.  For this reason, the HO$_{\textit{br}}$ levels in (b) have less weight on the benzene ring compared to (c,d).  Despite the differences in reduction of the substrate between the systems depicted in  Figures~\ref{fgr:OrbitalSpectra}(b,c,d,e), the HO$_{\textit{br}}$ levels are surprisingly consistent in energy.

As can be seen in Figure~\ref{Structures:fgr}(a), for \sfrac{1}{2}D \sfrac{1}{2}D, the HO$_{\textit{br}}$ groups are on every other O$_{\textit{br}}$, whereas in Figure~\ref{Structures:fgr}(b,c), the HO$_{\textit{br}}$--O bonds for D \sfrac{1}{2}D and D D are on adjacent O$_{\textit{br}}$.  For this reason, the HO$_{\textit{br}}$ levels are essentially isoenergetic in Figure~\ref{fgr:OrbitalSpectra}(c,d,e), whereas in (b) the HO$_{\textit{br}}$ level is more stable by $\sim0.3$ eV.  This is consistent with the observed downshift by 0.1 eV of the HO$_{\textit{br}}$ surface levels upon reducing the coverage from 1 ML to \sfrac{1}{2} ML H@O$_{\textit{br}}$ on TiO$_2$(110) \cite{MiganiH2OJCTC2014}.

A peak at $\sim 12$ eV below the experimental Fermi level indicates the presence of intermolecular OH--O hydrogen bonding within the overlayer.  On the one hand, this may be used to fingerprint the presence of catechol which is not fully deprotonated. On the other hand, its absence suggests the catechol overlayer is fully dissociated.  These levels have significant $\sigma$-bonding character along the OH--O intermolecular hydrogen bond.  This is combined with benzene skeleton $\sigma$ orbitals.  For the D \sfrac{1}{2}D overlayer, the OH--O orbital has very little weight on the fully dissociated catechol molecule, as seen in Figure~\ref{fgr:OrbitalSpectra}(c).  In other words, the OH--O levels are mostly associated with \sfrac{1}{2}D catechol molecules.


\begin{figure}[!t]
\includegraphics[width=\columnwidth]{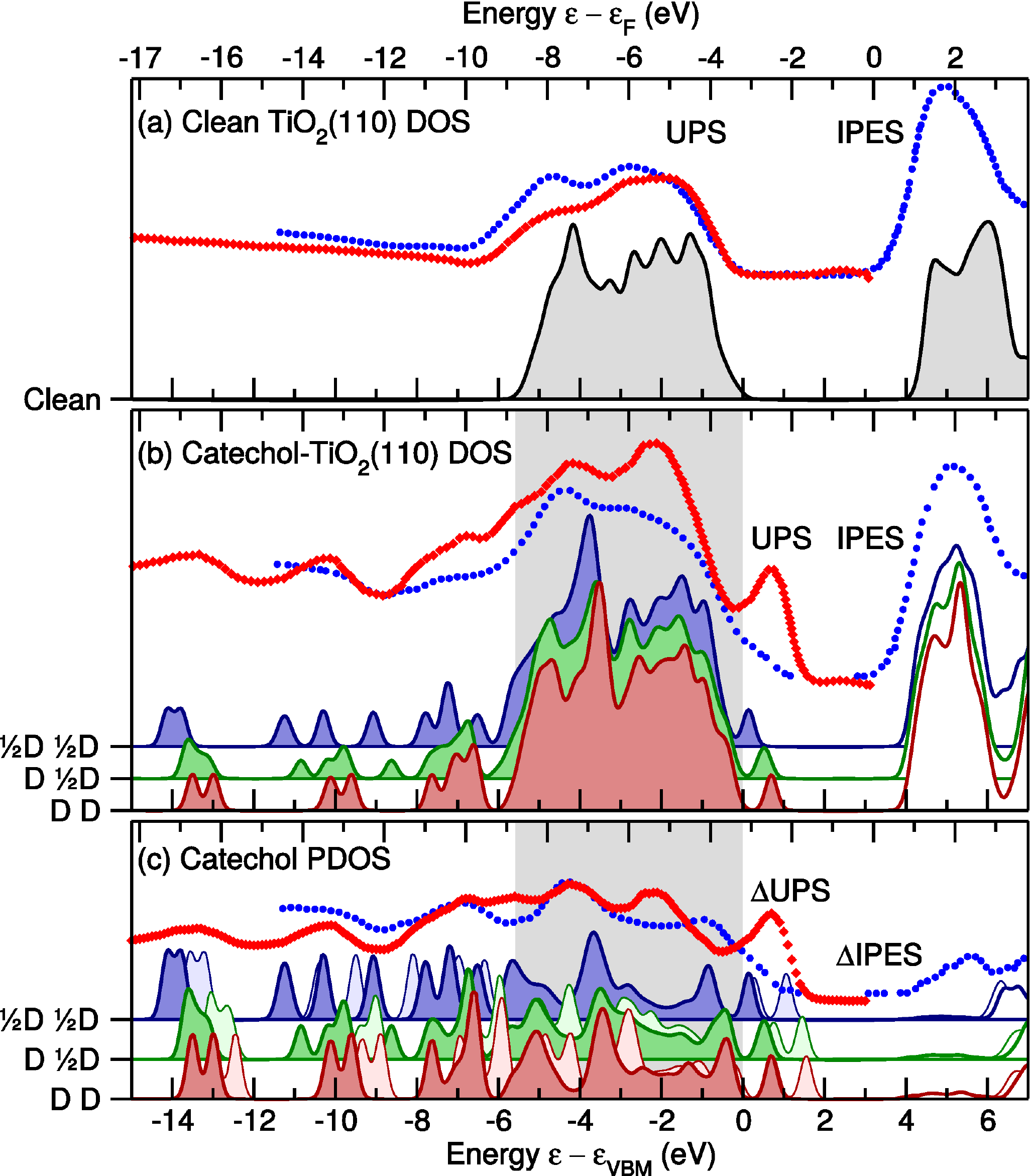}
\caption{$G_0W_0$ total DOS for (a) clean TiO$_2$(110) and (b) 1 ML catechol covered TiO$_2$(110), and (c) $G_0W_0$ (thick lines) and HSE DFT (thin lines) catechol PDOS, for half (\sfrac{1}{2}D \sfrac{1}{2}D, blue), mixed (D \sfrac{1}{2}D, green), and fully (D D, red) dissociated catechol overlayers, as compared to the experimental UPS and IPES from  ref\old{.}~\citenum{Rangan20104829} (blue circles) and UPS from ref\old{.}~\citenum{Diebold} (red diamonds) for (a) clean TiO$_2$(110), (b) catechol covered TiO$_2$(110), and (c) their difference spectra.  Energies are relative to the VBM  ($\varepsilon_{\textrm{VBM}}$) of clean TiO$_2$(110).  Filling denotes occupation.  }
\label{CatecholPDOS:fgr}
\noindent{\color{StyleColor}{\rule{\columnwidth}{1pt}}}
\end{figure}

In Figure~\ref{CatecholPDOS:fgr}, we compare the $G_0W_0$ DOS and PDOS for the clean, \sfrac{1}{2}D \sfrac{1}{2}D, D \sfrac{1}{2}D,  and D D catechol overlayers with IPES\cite{Rangan20104829} and UPS\cite{Diebold} for the (a) clean, (b) catechol covered, and their (c) difference spectra. 
In each case, there are several peaks outside the clean surface's VB region, shown in gray in Figure~\ref{CatecholPDOS:fgr}.  These are the peaks which are most easily distinguishable from the surface levels.  For overlayers which are highly hybridized with the surface, e.g., H$_2$O and H covered TiO$_2$(110) \cite{MiganiH2OJCTC2014,SunH2OAnatase}, it is difficult to disentangle surface and molecular levels using the difference between experimental spectra for the covered and clean surfaces, i.e., difference spectra.  Overall, our $G_0W_0$ DOS and PDOS agree qualitatively with the UPS/IPES and $\Delta$UPS/$\Delta$IPES spectra for catechol on TiO$_2$(110) from refs\old{.}~\citenum{Diebold} and \citenum{Rangan20104829}.  

Although there is a nice alignment between the computed and measured spectra for unoccupied levels (Figure~\ref{CatecholPDOS:fgr}(b)), these \new{substrate }levels are \old{highly }hybridized with the \old{surface}\new{catechol overlayer} (Figure~\ref{CatecholPDOS:fgr}(c)).  As a result, although the  unoccupied $G_0W_0$ DOS, PDOS, IPES and $\Delta$IPES levels agree, there is little to distinguish between the types of catechol overlayers.  This is not at all surprising, as these levels are predominantly Ti 3d in character, and should hybridize equally well with \sfrac{1}{2}D and D catechol.  However, there is a noticeable increase in PDOS intensity for the D D structure.  This is probably associated with a stronger coupling between O--Ti$_\textit{cus}$ compared to HO--Ti$_{\textit{cus}}$.  This results in a raising of the Ti$_{\textit{cus}}$ atoms out of the TiO$_2$(110) surface plane in Figure~\ref{Structures:fgr}(c) for the D D catechol overlayer.

By comparing the UPS peaks outside the VB region to the $G_0W_0$ DOS and PDOS, we \old{obtain clear identifying marks indicating a predominance of}\new{find features suggestive of} \sfrac{1}{2}D \sfrac{1}{2}D catechol in the UPS of ref~\citenum{Rangan20104829}, \old{while}\new{and D D catechol in} the UPS of ref~\citenum{Diebold}\old{ indicates a mostly D D catechol covered surface}.  \new{However, differences in detection setup and resolution between refs \citenum{Rangan20104829} and \citenum{Diebold} mean an absolute attribution of the measured spectra to \sfrac{1}{2}D \sfrac{1}{2}D and D D catechol, respectively, may be excessive.} \old{Specifically}\new{Nevertheless}, comparing the UPS catechol covered and difference spectra from ref~\citenum{Rangan20104829} with the $G_0W_0$ spectra, we find the following three fingerprints of the \sfrac{1}{2}D \sfrac{1}{2}D spectra. (1) A shoulder $\sim0.5$ eV above the VBM is \old{attributable to}\new{suggestive of} the \sfrac{1}{2}D \sfrac{1}{2} D catechol overlayer's HOMO. (2) A peak at $\sim-10$~eV is \old{attributable to}\new{suggestive of} HO$_{\textit{br}}$ interfacial levels. (3) A peak at $\sim-12$~eV is \old{attributable to}\new{suggestive of} intermolecular OH--O hydrogen bonds.  Performing a similar comparison to the catechol covered UPS from ref~\citenum{Diebold}, and the resulting difference spectra, we find the same three fingerprints, but of the D D spectra. (1) A well-separated peak $\sim0.8$ eV above the VBM is \old{attributable to}\new{suggestive of} the D D catechol overlayer's HOMO. (2) A more intense peak at $\sim-10$~eV is \old{attributable to}\new{suggestive of} HO$_{\textit{br}}$ interfacial levels. (3) A significant dip in the spectra at $\sim-12$~eV \old{indicates the absence of}\new{suggests fewer} intermolecular OH--O hydrogen bonds.  Note that, as HO$_{\textit{br}}$ levels are associated with the TiO$_2$(110) surface, this peak is absent from the PDOS, as the O$_{\textit{br}}$ atom is part of the surface.  

In Figure~\ref{CatecholPDOS:fgr}(c), we also compare catechol's PBE $G_0W_0$ and HSE DFT PDOS for \sfrac{1}{2}D \sfrac{1}{2}D, D \sfrac{1}{2}D, and D D overlayers on TiO$_2$(110).  In each case, the HSE DFT PDOS for occupied levels yields $\sim1$ eV weaker binding energies than PBE $G_0W_0$.  This upshift of the occupied molecular levels with HSE DFT is consistent with our previous results for the localized CH$_3$OH's a$^{_{''}}$ and H$_2$O's 3a$_1$ and 1b$_2$ levels on TiO$_2$(110) \cite{MiganiH2OJCTC2014,MiganiLong}. On the other hand, for the unoccupied levels, catechol's PDOS from HSE DFT and PBE $G_0W_0$ are consistent with each other, as the unoccupied molecular levels are highly hybridized with the substrate. This reinforces the finding that HSE fails to provide an accurate description of the interfacial level alignment for localized molecular levels\cite{MiganiH2OJCTC2014,MiganiLong}.  



\section{CONCLUSIONS}\label{Sect:Conclusions}

The structure of catechol overlayers on TiO$_2$(110) is characterized by a complex network of interfacial and intermolecular hydrogen bonds.  It is difficult to precisely establish the detailed structure of the catechol overlayer based solely on STM experiments. This is because simply deprotonating catechol's anchoring groups, while nearly isoenergetic,  strongly affects the position of the HOMO in this type II interface.  Hence, the extent of catechol's deprotonation on the surface determines the interface's photovoltaic efficiency.

We combine $G_0W_0$ level alignment with UPS measurements to identify the fingerprints 
of half, mixed, or fully dissociated catechol overlayers on TiO$_2$(110).  Only QP  techniques, such as $G_0W_0$, are sufficiently accurate to \old{attribute }robustly \new{predict} \old{the}\new{an} $\sim0.5$~eV energy difference between catechol's HOMO position in the \old{UPS spectra of refs\old{.}~\citenum{Rangan20104829} and \citenum{Diebold} to }half and fully dissociated catechol overlayers\old{, respectively}.  Moreover, besides the HOMO position, which indicates the extent of deprotonation of catechol's OH anchoring groups, the absence of a peak at $\sim -12$~eV \old{indicates}\new{is indicative of} a lack of intermolecular OH--O bonds.   Likewise, the presence of a peak at $\sim-10$~eV \old{indicates}\new{suggests} the presence of HO$_{\textit{br}}$ groups on the surface, which are formed upon deprotonation of catechol's OH anchoring groups.  

This distinct peak has also been observed for H$_2$O dissociated on bridging O vacancies, i.e., H$_2$O@O$_{\textit{br}}^{\textit{vac}}$, of a reduced TiO$_{2-x}$(110) surface \cite{DeSegovia,ThorntonH2ODissTiO2110,Krischok,MiganiH2OJCTC2014}.  
This peak has previously been considered a fingerprint of dissociated H$_2$O@O$_{\textit{br}}^{\textit{vac}}$.  However, we have shown this HO$_{\textit{br}}$ peak at $\sim-10$~eV may also arise from deprotonation of hydroxylated molecules on Ti$_{\textit{cus}}$.  Further, the position of the HO$_{\textit{br}}$ peak is rather insensitive to the degree of substrate reduction.  In UPS experiments, one should find this peak moves to slightly weaker binding energies as the coverage of HO$_{\textit{br}}$ is increased.  

While the energy of the unoccupied levels is rather insensitive to deprotonation of catechol's anchoring groups, the overlap of these \new{substrate }levels with those of the \old{substrate}\new{catechol overlayer} increases with deprotonation.  Combined with the destabilization of catechol's HOMO with deprotonation, this suggests fully deprotonated catechol overlayers should  have the greatest photovoltaic efficiency.  This work provides a road map for future studies of catechol's optical absorption based on the Bethe-Salpeter equation\cite{BetheSalpeterEqn,KresseBSE}, and catechol's subsequent rate of charge transport through the TiO$_2$ substrate using non-equilibrium Green's function methods\cite{NEGFMethodology,cnt_networks}.  

\section*{\large$\blacksquare$\normalsize\ ASSOCIATED CONTENT}
\subsubsection*{\includegraphics[height=8pt]{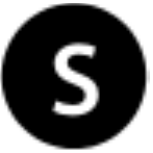} 
Supporting Information}
\noindent
The Supporting Information is available free of charge on the \href{http://pubs.acs.org}{ACS Publications website} at DOI: \href{http://dx.doi.org/10.1021/acs.jpcc.5605392}{10.1021/acs.jpcc.5605392}.

Comparison of $\rho(\textbf{r},\varepsilon_F+U)$ and $\int_{\varepsilon_F}^{\varepsilon_F+U}\rho(\textbf{r},\varepsilon)d\varepsilon$ line scans

(\href{http://pubs.acs.org/doi/suppl/10.1021/acs.jpcc.5b05392/suppl_file/jp5b05392_si_001.pdf}{PDF})

\section*{\large$\blacksquare$\normalsize\ AUTHOR INFORMATION}
\subsubsection*{Corresponding Author}
\noindent E-mail: \href{mailto:duncan.mowbray@gmail.com}{duncan.mowbray@gmail.com} (D.J.M.).

\noindent E-mail: \href{mailto:annapaola.migani@icn2.cat}{annapaola.migani@icn2.cat} (A.M.).
\subsubsection*{Notes} 
\noindent The authors declare no competing financial interest.
\section*{\large$\blacksquare$\normalsize\ ACKNOWLEDGMENTS} 

We acknowledge financial support from Spanish Grants (FIS2012-37549-C05-02, FIS2013-46159-C3-1-P, RYC-2011-09582, JCI-2010-08156); Generalitat de Catalunya (2014SGR301, XRQTC); Grupos Consolidados UPV/EHU del Gobierno Vasco (IT-578-13).
\bibliography{bibliography}



\end{document}